\documentclass[10pt,aps,prd,nofootinbib,showpacs,superscriptaddress,twocolumn,floatfix]{revtex4-2}

\usepackage{amsmath,amssymb}
\usepackage[utf8]{inputenc}
\usepackage{graphicx}
\usepackage{mathrsfs}
\usepackage{bm}
\usepackage{bbm}
\usepackage{mathtools}
\usepackage{indentfirst}
\usepackage{epstopdf}
\usepackage{color}
\usepackage{xcolor}
\usepackage{amssymb, amsmath,color, graphicx}
\usepackage[mathlines]{lineno}
\usepackage{booktabs}
\usepackage{braket}
\usepackage{placeins}
\usepackage{multirow}
\usepackage{slashed}
\usepackage{physics}
\usepackage{hyperref}
\newcommand{\diag}{\operatorname{diag}}
\newcommand{\bec}{\begin{center}}
\newcommand{\eec}{\end{center}}
\newcommand{\beq}{\begin{equation}}
\newcommand{\eeq}{\end{equation}}
\newcommand{\bea}{\begin{eqnarray}}
\newcommand{\eea}{\end{eqnarray}}

\begin{document}
\title{Chiral and $U(1)_A$ symmetries in background magnetic fields from lattice QCD}
\author{Heng-Tong Ding} %% Author name
\author{José Javier Hernández Hernández} %% Author name
\author{Dan Zhang} %% Author name
\affiliation{Key Laboratory of Quark and Lepton Physics (MOE) and Institute of
Particle Physics, Central China Normal University, Wuhan 430079, China}
\date{\today}
%------------------------------------------------------------------------------------
%       abstract
%------------------------------------------------------------------------------------
\begin{abstract}
We study chiral symmetry and singlet $U(1)_A$ symmetry in QCD in a background
magnetic field using lattice QCD.  We first clarify the neutral-sector
symmetry structure in a pure magnetic background, where the unequal
electric charges of the light quarks explicitly reduce the non-singlet
flavor symmetry.  We identify the neutral-pion--sigma susceptibility difference, $\chi_{\pi^0}-\chi_\sigma$, as the chiral-partner splitting associated with the surviving neutral non-singlet axial symmetry, and the neutral-pion--delta susceptibility difference, $\chi_{\pi^0}-\chi_{\delta^0}$, as the singlet $U(1)_A$ partner splitting. We also discuss the disconnected contribution
to the neutral-pion susceptibility and its continuum constraint.
Numerical results are obtained on fixed-scale $(2+1)$-flavor HISQ
ensembles with $m_l=m_s^{\rm phys}/10$, corresponding to a pion mass of
about $220~{\rm MeV}$ at vanishing magnetic field.  We find that the
neutral chiral-partner splitting increases with the magnetic field strength $eB$ at low temperature and decreases at sufficiently large $eB$ near the crossover, providing susceptibility-splitting counterparts of magnetic catalysis and inverse magnetic catalysis, respectively.  The singlet $U(1)_A$ partner splitting shows an analogous low-temperature enhancement and large-field suppression near the crossover, with the suppression setting in at larger $eB$  and remaining milder than in the chiral channel. These results provide a first lattice-QCD study of
neutral-sector probes of chiral and singlet $U(1)_A$ partner susceptibility splittings in background
magnetic fields.
\end{abstract}

\maketitle

%%%%%%%%%%%%%%%%%%%%%%%%%%%%%%%%%%%%%%%%%%%%%%%%%%%%%%%%%%%%
\section{Introduction}
\label{sec:intro}
%%%%%%%%%%%%%%%%%%%%%%%%%%%%%%%%%%%%%%%%%%%%%%%%%%%%%%%%%%%%
The fate of chiral symmetry and $U(1)_A$ symmetry is a central question
in finite-temperature QCD.  In the limit of two massless light quarks and in the absence of electromagnetism, the QCD action has the non-singlet chiral symmetry $SU(2)_L\times SU(2)_R$, in addition to the singlet vector symmetry $U(1)_V$.  The non-singlet chiral symmetry is
spontaneously broken in the low-temperature phase, and its breaking effects become strongly reduced
above the QCD crossover region.  The $U(1)_A$ symmetry is different:
it is broken by the quantum anomaly~\cite{Adler:1969gk,Bell:1969ts,tHooft:1976rip}, and its effective restoration at
high temperatures refers to the suppression of $U(1)_A$-breaking effects
in correlation functions rather than the disappearance of the anomaly
itself~\cite{Cohen:1996ng,Lee:1996zy,Shuryak:1993ee}.

The strength of $U(1)_A$ breaking near the chiral crossover is important
for several reasons.  In the two-flavor chiral limit it can affect the
possible order and universality class of the thermal transition~\cite{Pisarski:1983ms}, while at
finite quark masses it controls the splitting between scalar and
pseudoscalar channels that would otherwise become degenerate if
$U(1)_A$-breaking effects were sufficiently suppressed~\cite{Cohen:1996ng,Lee:1996zy}.  It is also
closely related to the infrared Dirac eigenvalue spectrum and to
topological fluctuations~\cite{Banks:1979yr,Ding:2023oxy,Leutwyler:1992yt,Evans:1996wf,Ding:2020xlj}.  At zero magnetic field, these questions have
been studied through scalar and pseudoscalar susceptibilities, in
particular by probing the degeneracy or splitting of chiral and $U(1)_A$
partner channels \cite{Aoki:2012yj,HotQCD:2012vvd,Buchoff:2013nra,
Ding:2020xlj}.

In a magnetic background, the role of the $U(1)_A$ anomaly is especially
interesting.  A pure magnetic field does not directly contribute to the axial anomaly, since the electromagnetic contribution is proportional to $F^{\rm em}\widetilde F^{\rm em}\propto{\bf E}\cdot{\bf B}$ and is
therefore absent for ${\bf E}=0$. 
This absence, however, does not mean that the magnetic field is irrelevant for axial $U(1)$ breaking. The gluonic anomaly remains in the divergence of the singlet axial current, while the external magnetic field can change the gauge-field ensemble on which this anomaly is realized. In particular, it can modify the topological fluctuations~\cite{Brandt:2024gso,Adhikari:2021lbl} and the infrared Dirac spectrum, thereby changing the magnitude of $U(1)_A$-breaking effects in
thermal observables. The magnetic-field dependence of
$U(1)_A$ partner splittings is therefore a nontrivial first-principles question.  This question is also conceptually connected to anomalous phenomena in strong magnetic
fields, such as the chiral magnetic effect, where topology, axial-charge
fluctuations and magnetic fields are closely intertwined~\cite{Kharzeev:2007jp,Fukushima:2008xe,Kharzeev:2015znc}.

The magnetic-field dependence of chiral symmetry has been extensively
studied in lattice QCD~\cite{Yamamoto:2021oys,Endrodi:2024cqn,Brandt:2026dcd,Ding:2026gao}, mainly through the chiral condensate and related observables.  At low temperature, a magnetic field enhances the
chiral condensate, a phenomenon known as magnetic catalysis.  Around the
crossover region, however, lattice QCD calculations found the opposite behavior: the
light-quark condensate and the pseudo-critical temperature decrease with
increasing magnetic field.  This inverse magnetic catalysis shows
that the magnetic response of QCD matter is strongly temperature
dependent and that the crossover region is particularly sensitive to the
interplay between valence and sea-quark effects~\cite{DElia:2011koc, Bali:2011qj,Bali:2012zg,Bruckmann:2013oba,DElia:2018xwo,Ding:2022tqn,Ding:2026qzu}.

Much less is known from first principles about the magnetic-field
dependence of $U(1)_A$-breaking observables.  A recent $(2+1)$-flavor
Nambu--Jona-Lasinio model study, including anomalous magnetic moments of
quarks, suggested an inverse-magnetic-catalysis-like behavior in a susceptibility difference associated with axial $U(1)_A$ breaking and referred to it as axial
inverse magnetic catalysis~\cite{Wang:2021dcy}.  This provides an
interesting model-motivated scenario for how $U(1)_A$ breaking may respond to a magnetic field. A first-principles QCD study of such a magnetic response must therefore start from the symmetry structure in a magnetic background, which fixes the relevant axial partner channels. 

The key point is that the electric-charge matrix is not proportional to
the identity in light-flavor space.  Therefore, even for degenerate
light-quark masses, $m_u=m_d$, a magnetic field distinguishes the $u$
and $d$ quarks through their different electric charges.  As a result,
the full non-singlet flavor symmetry is explicitly reduced, and the
charged and neutral isovector channels are no longer symmetry
equivalent.  This means that the standard zero-field partner-channel pattern for chiral
and $U(1)_A$ symmetry restoration cannot be used at nonzero magnetic
field without first reexamining the symmetry structure.  
This leads to a neutral-sector classification of the corresponding chiral and axial $U(1)$ susceptibility splittings, which, to our knowledge, has not been carried out before in a magnetic background.

In this work we perform this analysis and show that a clean symmetry classification can be obtained for the neutral sector.  In particular, the
difference between the neutral-pion susceptibility and the
scalar-isoscalar susceptibility, $\chi_{\pi^0}-\chi_\sigma$, is the chiral-partner splitting associated with the surviving neutral non-singlet axial symmetry, while the difference between the
neutral-pion susceptibility and the neutral scalar-isovector
susceptibility, $\chi_{\pi^0}-\chi_{\delta^0}$, is the corresponding axial
$U(1)_A$ partner splitting.  We then present lattice QCD results for
these observables in background magnetic fields using fixed-scale
$(2+1)$-flavor HISQ ensembles with $m_l=m_s^{\rm phys}/10$, for which
the Goldstone pion mass is about $220~{\rm MeV}$ at $eB=0$.  The
fixed-scale setup allows for a direct comparison of different temperatures
at the same physical value of $eB$, since a fixed flux quantum corresponds to the same magnetic-field strength when the lattice spacing is held fixed.

The paper is organized as follows. In
Sec.~\ref{sec:theory_observables} we discuss the symmetry structure of
QCD in a pure magnetic background and identify the neutral-channel probes
of chiral and $U(1)_A$ symmetries. In Sec.~\ref{sec:lattice_setup} we
describe the lattice setup and the construction of the measured
observables. Numerical results are presented in Sec.~\ref{sec:results},
followed by conclusions in Sec.~\ref{sec:conclusions}.

%%%%%%%%%%%%%%%%%%%%%%%%%%%%%%%%%%%%%%%%%%%%%%%%%%%%%%%%%%%%
\section{Symmetry structure and observables in a magnetic background}
\label{sec:theory_observables}
%%%%%%%%%%%%%%%%%%%%%%%%%%%%%%%%%%%%%%%%%%%%%%%%%%%%%%%%%%%%

We consider the light-quark sector of QCD in a classical external
electromagnetic background.  For the light doublet
\begin{equation}
  \psi_l=\begin{pmatrix}u\\ d\end{pmatrix},
\end{equation}
the continuum covariant derivative is
\begin{equation}
\begin{split}
  D_\mu={}&\partial_\mu+i g A^a_\mu t^a \\
           &+i e A^{\rm em}_\mu Q .
\end{split}
\label{eq:covD}
\end{equation}
The electric-charge matrix is
\begin{equation}
  Q=\diag\left(\frac23,-\frac13\right)
   =\frac{1}{6}{\bf 1}+\frac{1}{2}\tau^3 .
\label{eq:charge_matrix}
\end{equation}
Here, $Q$ acts in flavor space and should not be confused with the
topological charge. The matrix $\tau^3=\diag(1,-1)$ is the third Pauli
matrix in the $(u,d)$ flavor space. It distinguishes the neutral
isospin direction, whereas $\tau^1$ and $\tau^2$ mix the $u$ and $d$
flavors.

Let us first recall the zero-field symmetry structure. In the absence
of electromagnetism and in the two-flavor chiral limit, the classical
light-quark action is invariant under independent rotations of the
left- and right-handed fields,
\begin{equation}
  \psi_L\to U_L\psi_L,
  \qquad
  \psi_R\to U_R\psi_R,
  \qquad U_{L,R}\in U(2).
\end{equation}
Equivalently, the symmetry may be written as
\begin{equation}
  SU(2)_L\times SU(2)_R\times U(1)_V\times U(1)_A .
\end{equation}
The singlet vector symmetry $U(1)_V$ is the common phase rotation of all
quarks and is associated with baryon-number conservation. The
non-singlet chiral group $SU(2)_L\times SU(2)_R$ can be equivalently
described in terms of vector and axial directions. A vector rotation
acts in the same way on left- and right-handed fields,
\begin{equation}
  \psi_L\to V\psi_L,
  \qquad
  \psi_R\to V\psi_R,
  \qquad V\in SU(2),
\end{equation}
and forms the usual isospin symmetry $SU(2)_V$. An axial rotation acts
oppositely on left- and right-handed fields,
\begin{equation}
  \psi_L\to A^{-1}\psi_L,
  \qquad
  \psi_R\to A\psi_R,
  \qquad A\in SU(2),
\end{equation}
and corresponds to the non-singlet axial directions, denoted here by
$SU(2)_A$. Equivalently, in Dirac notation the vector generators are
$\tau^a$, while the axial generators are $\tau^a\gamma_5$. Thus
$SU(2)_V$ and $SU(2)_A$ are not additional independent groups beyond
$SU(2)_L\times SU(2)_R$; they are the vector and axial directions of the
same non-singlet chiral symmetry.

The QCD vacuum preserves the vector isospin symmetry $SU(2)_V$, but
spontaneously breaks the non-singlet axial symmetry $SU(2)_A$. Thus,
when one speaks of chiral symmetry restoration in two-flavor QCD, one
mainly refers to the restoration of the $SU(2)_A$ directions, or
equivalently to the restoration of the full
$SU(2)_L\times SU(2)_R$ symmetry up to the unbroken vector subgroup. The singlet axial symmetry $U(1)_A$ is different: it is broken by the
anomaly even in the chiral limit.

A nonzero magnetic field changes this classification already at the
level of the classical kinetic term. A flavor generator $T$ remains a
symmetry only if it commutes with the electromagnetic charge matrix,
\begin{equation}
  [T,Q]=0 .
\label{eq:commutator_condition}
\end{equation}
Since $Q$ is diagonal in the standard $(u,d)$ basis,
\begin{equation}
\begin{gathered}
  [{\bf 1},Q]=[\tau^3,Q]=0,\\
  [\tau^1,Q]\ne0,\qquad [\tau^2,Q]\ne0 .
\end{gathered}
\label{eq:commutators_Q}
\end{equation}
Thus, even for degenerate bare light-quark masses, $m_u=m_d\equiv m_l$,
the magnetic field explicitly distinguishes the $u$ and $d$ flavors
through their different electric charges. Consequently,
\begin{equation}
\begin{gathered}
  SU(2)_V \longrightarrow U(1)_V^{(3)},\\
  SU(2)_A \longrightarrow U(1)_A^{(3)} .
\end{gathered}
\label{eq:symmetry_reduction_B}
\end{equation}
Here $U(1)_V^{(3)}$ denotes the neutral vector subgroup generated by
$\tau^3$. Explicitly,
\begin{equation}
  \psi_l\to e^{i\alpha\tau^3/2}\psi_l,
  \qquad
  u\to e^{i\alpha/2}u,
  \quad
  d\to e^{-i\alpha/2}d .
\end{equation}
This corresponds to the conservation of the third component of isospin,
or equivalently the difference between $u$- and $d$-quark numbers.  It
should not be confused with the singlet $U(1)_V$ baryon-number symmetry,
which remains present separately.  Similarly, $U(1)_A^{(3)}$ denotes the
neutral non-singlet axial direction generated by $\tau^3\gamma_5$.

The superscript ``3'' in $U(1)_V^{(3)}$ and $U(1)_A^{(3)}$ denotes the
third direction in isospin space, not the spatial direction of the
magnetic field.  It appears because the electric-charge matrix is
diagonal in the $(u,d)$ flavor basis and is proportional to a
combination of ${\bf 1}$ and $\tau^3$.  The spatial orientation of the
magnetic field affects rotational symmetry and the dynamics of charged
states, but it is not the origin of the $\tau^3$ flavor structure.

It is useful to state the anomaly equation explicitly~\cite{Adler:1969gk,Bell:1969ts,Fujikawa:1979ay}.  For an axial
current with flavor generator $T$,
\begin{equation}
  J^\mu_{5,T}=\bar\psi\gamma^\mu\gamma_5T\psi,
\end{equation}
the divergence has the schematic form
\begin{equation}
\begin{aligned}
  \partial_\mu J^\mu_{5,T}
  ={}& i\bar\psi\{M_q,T\}\gamma_5\psi \\
     &+2\,\tr_f(T)\,q(x) \\
     &+\frac{e^2}{8 \pi^2}N_c
       \tr_f(TQ^2)
       F^{\rm em}_{\mu\nu}
       \widetilde F^{{\rm em}\,\mu\nu}.
\end{aligned}
\label{eq:general_axial_anomaly}
\end{equation}
Here $M_q=\left(\begin{smallmatrix}m_u&0\\0&m_d\end{smallmatrix}\right)$
is the light-quark mass matrix,
\begin{equation}
  q(x)=\frac{g^2}{32\pi^2}
  G^a_{\mu\nu}\widetilde G^{a\mu\nu}
\end{equation}
is the gluonic topological charge density, and $\tr_f$ denotes a trace
in flavor space.  The first term in \autoref{eq:general_axial_anomaly}
is the explicit breaking by quark masses.  The second term is the
gluonic anomaly~\cite{tHooft:1976rip}.  It is present for the singlet generator $T={\bf 1}$,
because $\tr_f({\bf 1})=N_f$, but it vanishes for non-singlet generators
such as $T=\tau^a$, because $\tr_f(\tau^a)=0$.

The last term in \autoref{eq:general_axial_anomaly} is the
electromagnetic contribution to the axial anomaly.  It can distinguish
different flavor generators through the factor $\tr_f(TQ^2)$.  However,
for the pure magnetic background considered here,
\begin{equation}
  F^{\rm em}_{\mu\nu}
  \widetilde F^{{\rm em}\,\mu\nu}
  \propto {\bf E}\cdot{\bf B}=0,
\end{equation}
because ${\bf E}=0$.  Therefore the electromagnetic anomaly
contribution is absent in the present setup.  The gluonic singlet axial
anomaly remains present, and a magnetic field can still modify the size
of $U(1)_A$-breaking effects in correlation functions.  Thus the effective restoration of the singlet 
$U(1)_A$  refers to the suppression of anomaly-induced $U(1)_A$-breaking
effects in suitable observables, not to the disappearance of the anomaly
equation. 

%%%%%%%%%%%%%%%%%%%%%%%%%%%%%%%%%%%%%%%%%%%%%%%%%%%%%%%%%%%%
\subsection{Neutral scalar and pseudoscalar channels}
\label{subsec:neutral_operators}
%%%%%%%%%%%%%%%%%%%%%%%%%%%%%%%%%%%%%%%%%%%%%%%%%%%%%%%%%%%%

We define the flavor-diagonal scalar and pseudoscalar bilinears as
\begin{equation}
  S_f=\bar f f,
  \qquad
  P_f=i\bar f\gamma_5 f,
  \qquad f=u,d .
\end{equation}
The neutral pseudoscalar and scalar operators are
\begin{equation}
\begin{gathered}
  \pi^0=\frac{P_u-P_d}{\sqrt2},
  \qquad
  \eta=\frac{P_u+P_d}{\sqrt2},\\[1mm]
  \delta^0=\frac{S_u-S_d}{\sqrt2},
  \qquad
  \sigma=\frac{S_u+S_d}{\sqrt2} .
\end{gathered}
\label{eq:neutral_operators}
\end{equation}
Here $\pi^0$ and $\delta^0$ are the neutral components of the
isovector pseudoscalar and scalar multiplets, while $\eta$ and $\sigma$
are isoscalar pseudoscalar and scalar operators.

At zero magnetic field, exact isospin symmetry makes the three isovector components equivalent. We denote the full isovector pseudoscalar and scalar multiplets by $\pi$ and
$\delta$, respectively. The standard singlet-axial pairings are~\cite{HotQCD:2012vvd,Buchoff:2013nra}
\begin{equation}
  U(1)_A^{\rm singlet}:
  \quad \pi\leftrightarrow\delta,
  \quad \sigma\leftrightarrow\eta .
\label{eq:pairing_U1A_B0}
\end{equation}
Hereafter, $U(1)_A^{\rm singlet}$ denotes the usual anomalous
axial $U(1)_A$ discussed above; the superscript ``singlet'' is used only
to distinguish it from the neutral non-singlet axial direction introduced
below. The corresponding non-singlet chiral pairings are~\cite{HotQCD:2012vvd,Buchoff:2013nra}
\begin{equation}
  SU(2)_A:
  \quad \pi\leftrightarrow\sigma,
  \quad \delta\leftrightarrow\eta .
\label{eq:pairing_SU2A_B0}
\end{equation}

At nonzero magnetic field, the charged directions are no longer
symmetry equivalent to the neutral direction. 
In the neutral sector, the singlet axial rotation still gives
\begin{equation}
U(1)_A^{\rm singlet}:
\quad
\pi^0\leftrightarrow\delta^0,
\qquad
\sigma\leftrightarrow\eta .
\label{eq}
\end{equation}
Therefore the neutral-pion--delta susceptibility difference $\chi_{\pi^0}-\chi_{\delta^0}$
is the singlet $U(1)_A$ partner splitting used in this work.~\footnote{The
complementary splitting $\chi_\sigma-\chi_\eta$ is another singlet-axial
partner difference. We focus on $\chi_{\pi^0}-\chi_{\delta^0}$ because
the full integrated $\chi_{\pi^0}$ is fixed by the non-singlet axial
Ward identity, whereas the singlet pseudoscalar channel requires noisy
disconnected pseudoscalar loops. This choice also permits a direct
comparison with the neutral chiral-partner splitting
$\chi_{\pi^0}-\chi_\sigma$, which shares the same neutral-pion
susceptibility.}

On the other hand, the surviving neutral non-singlet axial
transformation is generated by $\tau^3\gamma_5$.  We denote this
remaining chiral axial direction by $U(1)_A^{(3)}$, where the
superscript ``3'' labels the neutral isospin direction and distinguishes it
from the singlet axial $U(1)_A^{\rm singlet}$.  It gives
\begin{equation}
  U(1)_A^{(3)}:
  \quad \pi^0\leftrightarrow\sigma,
  \quad \delta^0\leftrightarrow\eta .
\label{eq:neutral_chiral_pairing_B}
\end{equation}
Thus, the neutral-pion--sigma susceptibility difference at nonzero
magnetic field,  
$\chi_{\pi^0}-\chi_\sigma$, is the chiral-partner splitting associated with the surviving neutral
non-singlet axial symmetry.

The charged channels $\pi^\pm$ and $\delta^\pm$ are not used as the
primary symmetry probes here, because they are associated with flavor
directions that do not commute with $Q$ and also couple directly to the
magnetic field through Landau-level dynamics.

%%%%%%%%%%%%%%%%%%%%%%%%%%%%%%%%%%%%%%%%%%%%%%%%%%%%%%%%%%%%
\subsection{Connected and disconnected susceptibilities}
\label{subsec:connected_disconnected}
%%%%%%%%%%%%%%%%%%%%%%%%%%%%%%%%%%%%%%%%%%%%%%%%%%%%%%%%%%%%

The symmetry probes identified above are susceptibilities of the neutral
mesonic operators ${\cal O}=\pi^0,\eta,\delta^0,\sigma$.  For each
operator we define
\begin{equation}
  \chi_{\cal O}
  =
  \sum_x
  \left[
    \langle{\cal O}(x){\cal O}(0)\rangle
    -
    \langle{\cal O}\rangle^2
  \right].
\label{eq:susc_definition}
\end{equation}
The subtraction removes the constant vacuum piece.  It is essential in
scalar channels, where the one-point function is generally nonzero,
whereas it vanishes for pseudoscalar channels at $\theta=0$ by parity in a pure magnetic background.

The Wick contractions of \autoref{eq:susc_definition} naturally
separate into connected and disconnected contributions.  Let $m_f$ be
the quark mass of flavor $f$, and let $D_f$ be the corresponding
massless Dirac operator, including the coupling to the magnetic field.
We denote
\begin{equation}
  {\cal M}_f=D_f+m_f,
  \qquad
  G_f(x,y)={\cal M}_f^{-1}(x,y) ,
\label{eq:fermion_matrix_def}
\end{equation}
where $G_f$ is the quark propagator in a fixed gauge and electromagnetic
background.

The connected pseudoscalar and scalar susceptibilities entering the
neutral channels are defined as
\begin{equation}
\begin{aligned}
  \chi_{5,\rm con}
  &=\frac{1}{2}\sum_{f=u,d}\sum_x
  \left\langle
  {\rm tr}\,\gamma_5G_f(x,0)\gamma_5G_f(0,x)
  \right\rangle,\\
  \chi_{\rm con}
  &=-\frac{1}{2}\sum_{f=u,d}\sum_x
  \left\langle
  {\rm tr}\,G_f(x,0)G_f(0,x)
  \right\rangle .
\end{aligned}
\label{eq:connected_susc_def}
\end{equation}
Here ${\rm tr}$ denotes a trace over color and Dirac indices.

The disconnected parts are expressed in terms of the local scalar and
pseudoscalar loops,
\begin{equation}
  L_f(x)={\rm tr}\,G_f(x,x),
  \qquad
  L_5^f(x)={\rm tr}\,\gamma_5G_f(x,x).
\label{eq:local_loops}
\end{equation}
For the flavor combinations relevant to the neutral sector, define
\begin{equation}
  L_{5,\pm}=L_5^u\pm L_5^d,
  \qquad
  L_\pm=L^u\pm L^d .
\end{equation}
Then
\begin{equation}
  \chi^{(\pm)}_{5,\rm disc}
  =
  \frac12\sum_x
  \Big[
  \langle L_{5,\pm}(x)L_{5,\pm}(0)\rangle
  -\langle L_{5,\pm}\rangle^2
  \Big],
\label{eq:pseudo_disc_pm_def}
\end{equation}
and
\begin{equation}
  \chi^{(\pm)}_{\rm disc}
  =
  \frac12\sum_x
  \Big[
  \langle L_\pm(x)L_\pm(0)\rangle
  -\langle L_\pm\rangle^2
  \Big].
\label{eq:scalar_disc_pm_def}
\end{equation}

With these definitions, the neutral susceptibilities decompose as
\begin{equation}
\begin{aligned}
  \chi_{\pi^0}
    &=\chi_{5,\rm con}-\chi^{(-)}_{5,\rm disc} ,\\
  \chi_{\eta}
    &=\chi_{5,\rm con}-\chi^{(+)}_{5,\rm disc} ,\\
  \chi_{\delta^0}
    &=\chi_{\rm con}+\chi^{(-)}_{\rm disc} ,\\
  \chi_\sigma
    &=\chi_{\rm con}+\chi^{(+)}_{\rm disc} .
\end{aligned}
\label{eq:neutral_susc_decomp}
\end{equation}
This decomposition connects the symmetry probes directly to the
connected and disconnected Wick contractions.

For the scalar sector it is useful to introduce the full trace
\begin{equation}
  X_f={\rm Tr}\,{\cal M}_f^{-1},
\end{equation}
where now the trace includes color, Dirac, and spacetime indices.
Note that \autoref{eq:scalar_disc_pm_def} can be written equivalently as
\begin{equation}
  \chi^{(\pm)}_{\rm disc}
  =
  \frac{1}{2V_4}
  \Big[
  \langle (X_u\pm X_d)^2\rangle
  -\langle X_u\pm X_d\rangle^2
  \Big] .
\label{eq:lattice_scalar_disc_pm}
\end{equation}
Thus
\begin{equation}
\begin{split}
  \chi^{(+)}_{\rm disc}
  -\chi^{(-)}_{\rm disc}
  =
  \frac{2}{V_4}
  \Big[
  \langle X_uX_d\rangle
  -\langle X_u\rangle\langle X_d\rangle
  \Big] .
\end{split}
\label{eq:chidisc_plus_minus_difference}
\end{equation}
This difference measures the covariance between the full $u$- and
$d$-quark scalar traces.  It becomes useful at nonzero magnetic field
because the light quark masses remain degenerate, $m_u=m_d$, while the
corresponding fermion matrices differ, ${\cal M}_u\ne{\cal M}_d$, due to
$q_u\ne q_d$.

Finally, using \autoref{eq:neutral_susc_decomp}, 
the neutral chiral splittings are
\begin{align}
  \chi_{\pi^0}-\chi_\sigma
  &=
  \Delta_{\rm con}
  -\chi^{(-)}_{5,\rm disc}
  -\chi^{(+)}_{\rm disc},
  \label{eq:pi_sigma_decomp}
  \\
  \chi_\eta-\chi_{\delta^0}
  &=
  \Delta_{\rm con}
  -\chi^{(+)}_{5,\rm disc}
  -\chi^{(-)}_{\rm disc}.
  \label{eq:eta_delta_decomp}
\end{align}
These two differences correspond to the neutral chiral-partner pairs
$\pi^0\leftrightarrow\sigma$ and $\eta\leftrightarrow\delta^0$. Here $\Delta_{\rm con}\equiv \chi_{5,\rm con}-\chi_{\rm con}$ is the connected scalar--pseudoscalar splitting.
The neutral
singlet-axial splittings are 
\begin{align}
  \chi_{\pi^0}-\chi_{\delta^0}
  &=
  \Delta_{\rm con}
  -\chi^{(-)}_{5,\rm disc}
  -\chi^{(-)}_{\rm disc},
  \label{eq:pi_delta_decomp}
  \\
  \chi_\eta-\chi_\sigma
  &=
  \Delta_{\rm con}
  -\chi^{(+)}_{5,\rm disc}
  -\chi^{(+)}_{\rm disc}.
  \label{eq:eta_sigma_decomp}
\end{align}
 These two differences
correspond to the neutral-sector $U(1)_A$ partner pairs
$\pi^0\leftrightarrow\delta^0$ and $\eta\leftrightarrow\sigma$.

If the neutral chiral partner degeneracies are realized, one obtains
\begin{equation}
  \chi^{(+)}_{\rm disc}
  -
  \chi^{(-)}_{\rm disc}
  =
  \chi^{(+)}_{5,\rm disc}
  -
  \chi^{(-)}_{5,\rm disc}.
\label{eq:disc_relation_chiral_restoration}
\end{equation}
If the singlet axial $U(1)$ partner degeneracies are also exactly realized,~\autoref{eq:disc_relation_chiral_restoration} further implies
\begin{equation}
  \chi^{(+)}_{\rm disc}
  -
  \chi^{(-)}_{\rm disc}
  =
  \chi^{(+)}_{5,\rm disc}
  -
  \chi^{(-)}_{5,\rm disc}
  =0 .
\label{eq:disc_relation_chiral_u1_restoration}
\end{equation}
~\autoref{eq:disc_relation_chiral_u1_restoration} is therefore an idealized statement corresponding to simultaneous exact neutral-chiral and singlet-axial partner degeneracy. At finite nonzero light-quark
mass, effective restoration is generally characterized by approximate
partner degeneracy, so the combinations in~\autoref{eq:disc_relation_chiral_u1_restoration}  may be strongly suppressed but need not vanish identically. 

%%%%%%%%%%%%%%%%%%%%%%%%%%%%%%%%%%%%%%%%%%%%%%%%%%%%%%%%%%%%
\subsection{Neutral-pion susceptibility and its disconnected part}
\label{subsec:pi0_disc}
%%%%%%%%%%%%%%%%%%%%%%%%%%%%%%%%%%%%%%%%%%%%%%%%%%%%%%%%%%%%

The neutral-pion susceptibility enters both symmetry-sensitive
differences discussed above, $\chi_{\pi^0}-\chi_{\delta^0}$ and
$\chi_{\pi^0}-\chi_\sigma$.  It is therefore important to specify how
the full $\chi_{\pi^0}$ is obtained in practice.

At nonzero magnetic field the local $u$- and $d$-quark propagators are
different because $q_u\ne q_d$, even when $m_l=m_u=m_d$.  Therefore the
neutral-pion channel can contain a local disconnected contribution.
However, for the integrated susceptibility there is a useful continuum
constraint.  In the continuum, or in a lattice regularization with exact chiral symmetry, the integrated
pseudoscalar loop
\begin{equation}
  {\cal P}_f
  =
  {\rm Tr}\,\gamma_5{\cal M}_f^{-1}
\end{equation}
receives contributions only from exact zero modes and becomes an
index-type quantity~\cite{Fujikawa:1979ay,Leutwyler:1992yt},
\begin{equation}
  {\cal P}_f=\frac{{\rm index}(D_f)}{m_l}.
\end{equation}
In a pure magnetic background, the electromagnetic topological term is
absent because $F^{\rm em}\widetilde F^{\rm em}\propto
{\bf E}\cdot{\bf B}=0$.  The remaining index is therefore controlled by
the flavor-blind gluonic topological charge, implying
\begin{equation}
  {\cal P}_u={\cal P}_d
\end{equation}
configuration by configuration in the continuum pure-$B$ theory.  Thus
the integrated minus pseudoscalar disconnected susceptibility vanishes,
\begin{equation}
  \chi^{(-)}_{5,\rm disc}=0,
\end{equation}
as an integrated continuum statement.  This does not imply that the
local $u$- and $d$-pseudoscalar loops are equal point by point.

Direct measurements of the pseudoscalar disconnected susceptibilities
$\chi^{(\pm)}_{5,\rm disc}$ are numerically delicate.  They require
disconnected-loop measurements of
${\cal P}_f={\rm Tr}\,\gamma_5{\cal M}_f^{-1}$, which are statistically
expensive and are sensitive to topology and zero modes.  In addition,
for staggered fermions the $\gamma_5$ insertion and the associated taste structure have to be treated carefully~\cite{Kilcup:1986dg,Gregory:2007ev,Donald:2011if}. The plus combination
$\chi^{(+)}_{5,\rm disc}$ probes fluctuations of the singlet
pseudoscalar loop and is closely related to topological fluctuations,
whereas the minus combination $\chi^{(-)}_{5,\rm disc}$ is expected to
vanish in the continuum pure-$B$ theory\footnote{Using the continuum index relation above, one obtains
$\chi^{(+)}_{5,\rm disc}-\chi^{(-)}_{5,\rm disc}
=2\chi_{\rm top}/m_l^2$, where $\chi_{\rm top}$ is the
topological susceptibility. Hence, the exact vanishing in
\autoref{eq:disc_relation_chiral_u1_restoration} requires
$\chi_{\rm top}=0$ at fixed $m_l>0$. This is expected to occur
only asymptotically as $T\to\infty$; in the chiral limit, the
stronger condition $\chi_{\rm top}/m_l^2\to0$ is required.}.  Resolving the latter directly
therefore amounts to extracting a small residual signal from the
difference of two noisy quantities, and it can be sensitive to
finite-$a$ and taste-breaking effects.

For this reason, in the numerical analysis we obtain the full
neutral-pion susceptibility from the non-singlet axial Ward identity at both zero and nonzero magnetic fields~\cite{Ding:2020hxw},
\begin{equation}
  2m_l\,\chi_{\pi^0}
  =
  \langle\bar u u\rangle+\langle\bar d d\rangle .
\label{eq:pi0_ward_identity}
\end{equation}
This definition automatically gives the full $\chi_{\pi^0}$ needed in
both $\chi_{\pi^0}-\chi_{\delta^0}$ and $\chi_{\pi^0}-\chi_\sigma$.  If
$\chi_{5,\rm con}$ is also measured, the difference
$\chi_{\pi^0}-\chi_{5,\rm con}$ provides a useful check of the
finite-lattice-spacing size of $\chi^{(-)}_{5,\rm disc}$. 

%%%%%%%%%%%%%%%%%%%%%%%%%%%%%%%%%%%%%%%%%%%%%%%%%%%%%%%%%%%%
\section{Lattice setup}
\label{sec:lattice_setup}
%%%%%%%%%%%%%%%%%%%%%%%%%%%%%%%%%%%%%%%%%%%%%%%%%%%%%%%%%%%%

The numerical calculations are performed on the same gauge ensembles as
in Ref.~\cite{Ding:2021cwv}.  They use the tree-level improved
Symanzik gauge action and the highly improved staggered quark action for
$(2+1)$-flavor QCD in a background magnetic field.  The strange quark
mass is fixed to its physical value, while the degenerate light-quark
masses are set to \(m_u=m_d\equiv m_l=m_s^{\rm phys}/10\).  At
vanishing magnetic field this corresponds to
\(M_\pi=220.61(6)\,{\rm MeV}\) and \(f_K=112.50(2)\,{\rm MeV}\).
The lattice spacing is fixed to \(a\simeq0.117\,{\rm fm}\), or
\(a^{-1}\simeq1.685\,{\rm GeV}\), and the temperature is varied by
changing \(N_\tau\), with \(T=1/(aN_\tau)\).  All ensembles have spatial
volume \(32^3\).  The values \(N_\tau=6,8,10,12,14,16,18,24\), and 96
correspond to \(T\simeq281,211,169,140,120,105,94,70\), and
\(17\,{\rm MeV}\), respectively.

The magnetic field is introduced in the \(z\) direction by multiplying
the \(SU(3)\) gauge links entering the fermion matrix by flavor-dependent
\(U(1)\) phases.  With periodic boundary conditions the magnetic flux is
quantized as
\begin{equation}
  eB=\frac{6\pi N_b}{N_xN_y}\,a^{-2},
  \qquad N_b\in\mathbb Z .
\label{eq:eB_quantization}
\end{equation}
We use \(N_b=0,1,2,3,4,6,8,10,12,16,20,24,32,40\), and 48,
corresponding to magnetic fields from \(eB=0\) up to
\(eB\simeq2.51\,{\rm GeV}^2\).  For the largest flux,
\(N_b/N_\sigma^2\simeq5\%\).

Gauge configurations were generated with the rational hybrid Monte Carlo
algorithm and saved every five molecular-dynamics time units.  Depending
on \(N_\tau\) and \(eB\), the number of saved configurations ranges from
\(O(10^3)\) to \(O(10^4)\).  The present analysis uses the same saved
configurations as Ref.~\cite{Ding:2021cwv}.

For each configuration we measure the scalar traces entering
Eqs.~\eqref{eq:lattice_scalar_disc_pm} and
\eqref{eq:chidisc_plus_minus_difference}, together with
\(Y_f={\rm Tr}\,{\cal M}_f^{-2}\) for the connected scalar contribution.
We also measure the connected pion correlators and the light-quark
chiral condensates. In practice, the scalar connected susceptibility is evaluated from $Y_f$ as
\begin{align}
  \chi_{\rm con}=-\frac{1}{2V_4}\langle Y_u+Y_d\rangle .
\label{eq:lattice_measurement_chicon}
\end{align}
The scalar disconnected susceptibilities \(\chi_{\rm disc}^{(\pm)}\)
are evaluated using \autoref{eq:lattice_scalar_disc_pm}.  The neutral
scalar susceptibilities are then obtained from
\(\chi_{\delta^0}=\chi_{\rm con}+\chi^{(-)}_{\rm disc}\) and
\(\chi_\sigma=\chi_{\rm con}+\chi^{(+)}_{\rm disc}\).

The connected pion susceptibility \(\chi_{5,\rm con}\) is obtained from
the connected pion correlator, while the full neutral-pion susceptibility
is obtained from \autoref{eq:pi0_ward_identity} using the measured
chiral condensates~\footnote{We demonstrate that  $\chi^{(-)}_{5,\rm disc}\approx 0$ in our current lattice setup in Appendix~\ref{app:pi0_WI_check}.}. These quantities determine
\(\chi_{\pi^0}-\chi_{\delta^0}\) and \(\chi_{\pi^0}-\chi_\sigma\).

All results in this work are obtained at a single lattice spacing.  This
fixed-scale setup has the practical advantage that the magnetic-field
strength in physical units is fixed by the flux quantum \(N_b\) and does
not change when the temperature is varied by changing \(N_\tau\).  Thus
observables at different temperatures can be compared directly at the
same physical value of \(eB\), without an interpolation in the magnetic
field.  This is particularly useful for studying the temperature
dependence of the symmetry-sensitive observables at fixed \(eB\).

At the same time, since no continuum extrapolation is performed here, the
results should be interpreted within this fixed-scale setup.  The
continuum symmetry arguments discussed above guide the choice of
observables, while finite-\(a\) effects, including staggered taste
breaking and the absence of an exact continuum index theorem at finite
lattice spacing, should be kept in mind when interpreting the
neutral-pion disconnected sector.

%%%%%%%%%%%%%%%%%%%%%%%%%%%%%%%%%%%%%%%%%%%%%%%%%%%%%%%%%%%%
\section{Results}
\label{sec:results}
%%%%%%%%%%%%%%%%%%%%%%%%%%%%%%%%%%%%%%%%%%%%%%%%%%%%%%%%%%%%

In this section we present the temperature and magnetic-field dependence of the neutral-sector susceptibility splittings introduced above.  We focus on three dimensionless quantities,
\begin{align}
    O_A(B,T)
    &\equiv
    \frac{m_s^2}{f_K^4}
    \left[
    \chi_{\pi^0}(B,T)-\chi_{\delta^0}(B,T)
    \right],
    \label{eq:OA_def}
    \\
    O_\chi(B,T)
    &\equiv
    \frac{m_s^2}{f_K^4}
    \left[
    \chi_{\pi^0}(B,T)-\chi_{\sigma}(B,T)
    \right],
    \label{eq:Ochi_def}
    \\
    O_{\rm disc}(B,T)
    &\equiv
    \frac{m_s^2}{f_K^4}
    \left[
    \chi_{\rm disc}^{(+)}(B,T)
    -
    \chi_{\rm disc}^{(-)}(B,T)
    \right].
    \label{eq:Odisc_def}
\end{align}
Here $O_\chi$ is the neutral non-singlet chiral-partner
splitting associated with the surviving $U(1)^{(3)}_A$
symmetry, while $O_A$ is the singlet $U(1)_A$ partner
splitting in the neutral sector. With these definitions,
$O_A-O_\chi=O_{\rm disc}$, so $O_{\rm disc}$ displays the
difference between the two partner splittings. The prefactor $m_s^2/f_K^4$ makes the susceptibility differences dimensionless.  Since the scalar and pseudoscalar densities renormalize inversely to the quark mass, the combinations $m_s^2\chi$ are multiplicatively
renormalization-group invariant.  The same normalization is used for all temperatures and magnetic fields.  Since this work
is performed at a single lattice spacing, these observables should be regarded as fixed-scale dimensionless quantities.

\begin{figure*}[!htp]
  \centering
  \includegraphics[width=0.325\textwidth]{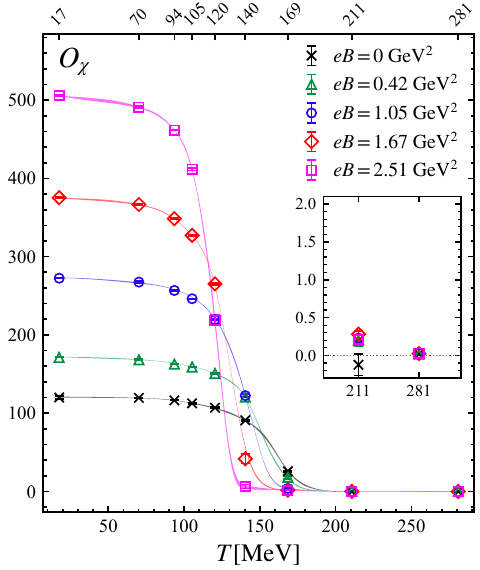}
  \includegraphics[width=0.325\textwidth]{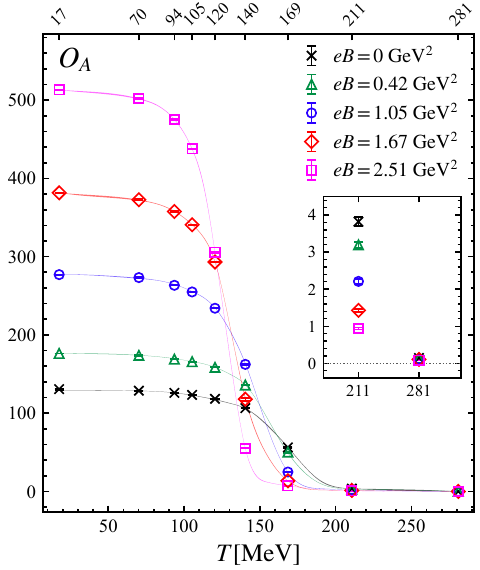}
  \includegraphics[width=0.325\textwidth]{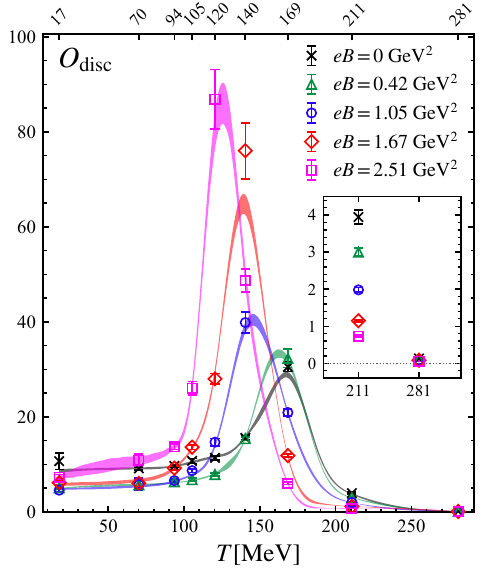}
\caption{Temperature dependence of the neutral-sector susceptibility splittings at fixed $eB$. From left to right: $O_\chi$, $O_A$, and $O_{\rm disc}=O_A-O_\chi$, defined in Eqs.~\eqref{eq:OA_def}--\eqref{eq:Odisc_def}. The insets magnify the high-temperature region.}
\label{fig:O_T_results}
\end{figure*}

\begin{figure*}[!htp]
  \centering
  \includegraphics[width=0.325\textwidth]{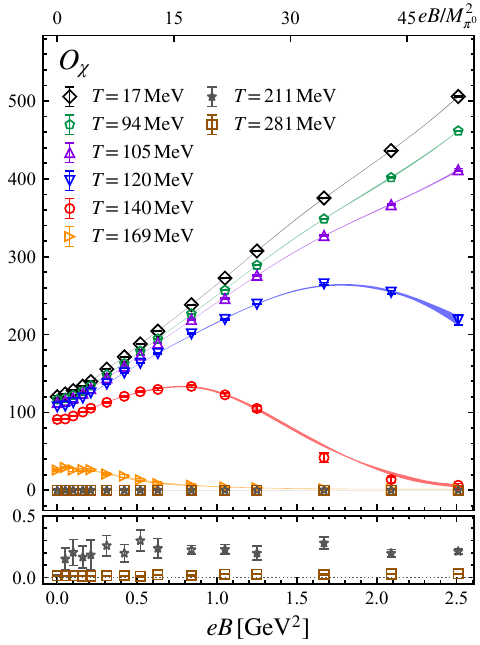}
  \includegraphics[width=0.325\textwidth]{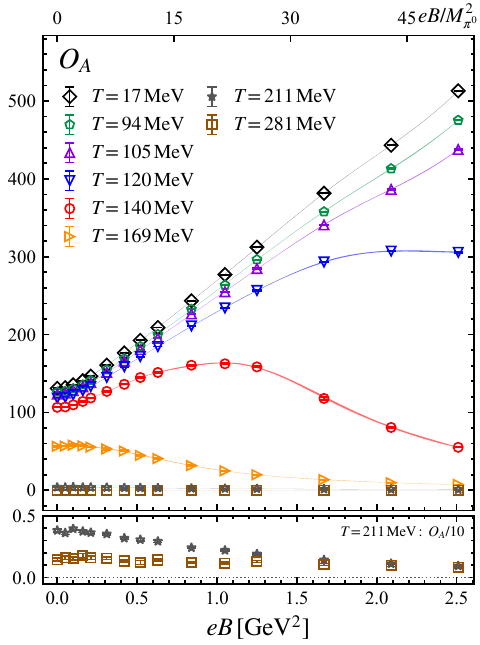}
  \includegraphics[width=0.325\textwidth]{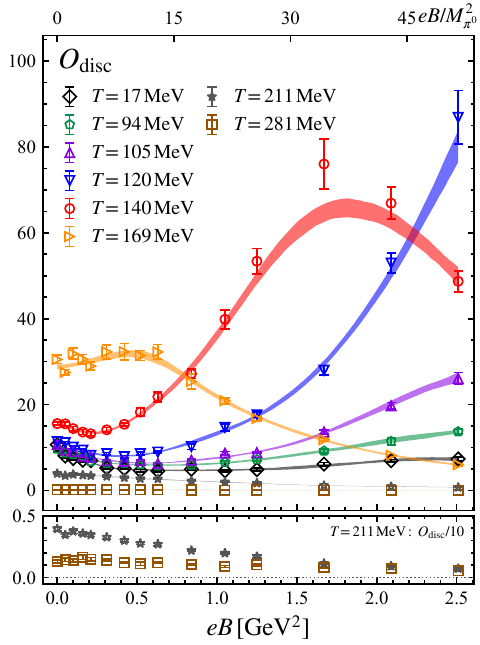}
\caption{Magnetic-field dependence of the same neutral-sector susceptibility splittings as shown in~\autoref{fig:O_T_results} at fixed temperature. From left to right: $O_\chi$, $O_A$, and $O_{\rm disc}$. The lower panels, where shown, magnify the high-temperature data; any rescaling is indicated in the plot. }
  \label{fig:O_eB_results}
\end{figure*}

We first discuss the temperature dependence of the neutral-sector susceptibility splittings, shown in \autoref{fig:O_T_results}.  The data points are the measured lattice results, while the shaded bands are shown only to guide the eye~\footnote{\autoref{fig:O_T_results} and \autoref{fig:O_eB_results} show complementary fixed-$eB$ and fixed-$T$ views of the same observables $O_i(T,eB)$. For each
observable and bootstrap sample, we perform a
covariance-weighted two-dimensional cubic B-spline
smoothing fit in the $(T,eB)$ plane. The shaded bands
represent the pointwise 16th--84th percentile ranges of
the resulting fitted surfaces evaluated along the slices
shown in the figures.}.  The left panel shows $O_\chi$, the neutral non-singlet chiral-partner splitting associated with the surviving $U(1)^{(3)}_A$ symmetry.  The middle panel shows $O_A$, the singlet $U(1)_A$ partner splitting in the neutral sector.  The right panel shows $O_{\rm disc}=O_A-O_\chi$, which displays the difference between these two partner splittings.

For the neutral non-singlet chiral splitting $O_\chi$, shown in the left panel of \autoref{fig:O_T_results}, the low-temperature data show a clear magnetic enhancement.  In particular, for $T\lesssim 105$ MeV, the data points at fixed temperature are  ordered by the magnetic-field strength, with larger $eB$
corresponding to a larger value of $O_\chi$.  This simple ordering starts to change when the temperature is raised to around $T=120$ MeV.  At this temperature and $T=140$ MeV, the magnetic-field dependence is no longer monotonic over the full range of $eB$: the data still show an enhancement at small and intermediate magnetic fields, while the largest magnetic fields lead to a reduced splitting. In other words, the low-temperature magnetic enhancement does not persist uniformly as $T$ is increased; the enhancement is reduced most strongly at the largest magnetic fields. We return to this point in the discussion of \autoref{fig:O_eB_results}, where the dependence on $eB$ is shown directly at fixed temperature.

At higher temperatures, $O_\chi$ is strongly suppressed.  At $T=169$ MeV it is already much smaller than in the low-temperature range, and at $T=211$ and $281$ MeV it is close to zero on the scale of the main figure.
The zoomed-in view of the two highest temperatures shows that only a very small residual signal can be resolved at $T=211$ MeV, while the splitting at
$T=281$ MeV is consistent with a vanishingly small value within the present resolution.  This behavior indicates the effective restoration of the surviving neutral non-singlet chiral symmetry, in the sense of the near-degeneracy of the neutral-pion and sigma partner channels at high temperature.

The singlet $U(1)_A$ partner splitting $O_A$, shown in the middle panel of \autoref{fig:O_T_results}, shows a similar low-temperature magnetic ordering as $O_\chi$.  For
$T\lesssim 120$ MeV, increasing $eB$ enhances $O_A$ at fixed temperature. As the temperature is raised from $T=120$ MeV to $T=169$ MeV, $O_A$ is strongly reduced.  Around $T=140$ MeV, the simple low-temperature ordering with $eB$ begins to change:
the enhancement remains visible at small and intermediate magnetic fields, while the largest magnetic fields give a noticeably smaller splitting.  This behavior parallels what is seen in $O_\chi$, but the two splittings should not be identified.

A key difference is that $O_A$ remains more persistent than $O_\chi$ at higher temperature.  In the intermediate-temperature range, and especially in the zoomed view around $T=211$ MeV, $O_A$ remains visibly nonzero, whereas $O_\chi$ is already strongly suppressed on the same scale.  This indicates
that the neutral non-singlet chiral partner degeneracy is reached earlier in temperature than the suppression of singlet $U(1)_A$ breaking effects.  At the
highest temperature, $T=281$ MeV, $O_A$ is strongly suppressed for all magnetic fields studied.  
We therefore focus here on the temperature dependence of the two partner splittings, and leave the detailed magnetic-field dependence of $O_A$ to~\autoref{fig:O_eB_results}.

The right panel of \autoref{fig:O_T_results} shows $O_{\rm disc}=O_A-O_\chi$.  Since this quantity is the difference between the singlet $U(1)$ partner splitting and the neutral non-singlet chiral splitting, it directly indicates where the two channels
separate.  The data show that $O_{\rm disc}$ is relatively small in the low-temperature region, develops a pronounced peak around the crossover region where $O_\chi$ changes most rapidly with $T$, and then becomes small again at high temperature.  This peak structure
shows that the difference between the two partner splittings is mainly localized near the crossover.  It also explains why the apparent restoration patterns of the two channels are not identical: the neutral non-singlet chiral
splitting is strongly suppressed already above the crossover, while the singlet $U(1)$ partner splitting persists to somewhat higher temperature.

We now turn to the magnetic-field dependence at fixed temperature, shown in~\autoref{fig:O_eB_results}. 
The left panel of \autoref{fig:O_eB_results} shows the neutral non-singlet chiral splitting $O_\chi$.  At low temperatures, namely $T=17$, $94$, and $105$ MeV, the data points increase monotonically with
$eB$ over the whole magnetic-field range studied.  This is the susceptibility counterpart of the magnetic enhancement of the non-singlet chiral sector. At
$T=120$ MeV, $O_\chi$ still increases with $eB$ up to
$eB\simeq 1.7~{\rm GeV}^2$, and then decreases at larger magnetic fields. Even at the largest magnetic field, however, the splitting remains above its $eB=0$ value at this temperature.
As the temperature is raised further, the turnover occurs at smaller magnetic field.  At $T=140$ MeV, $O_\chi$ reaches its maximum already around $eB\simeq 0.9~{\rm GeV}^2$ and then decreases rapidly with increasing $eB$.
At sufficiently large magnetic fields, the splitting becomes smaller than its $eB=0$ value.  At $T=169$ MeV, $O_\chi$ is already small and is further reduced as $eB$ is increased.  At $T=211$ and $281$ MeV, it remains close
to zero on the scale of the main plot.

\begin{figure*}[t]
  \centering
  \includegraphics[width=0.44\textwidth]{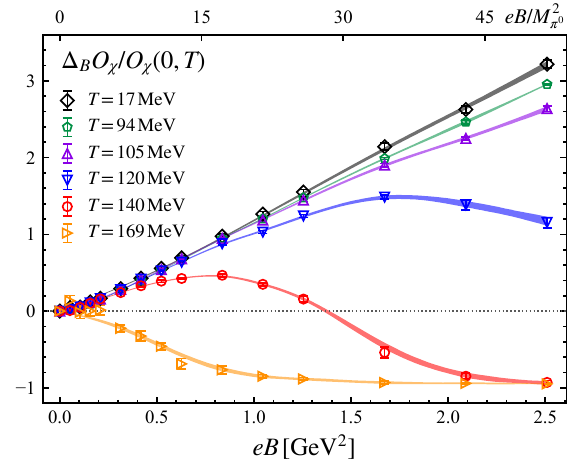}
  \includegraphics[width=0.44\textwidth]{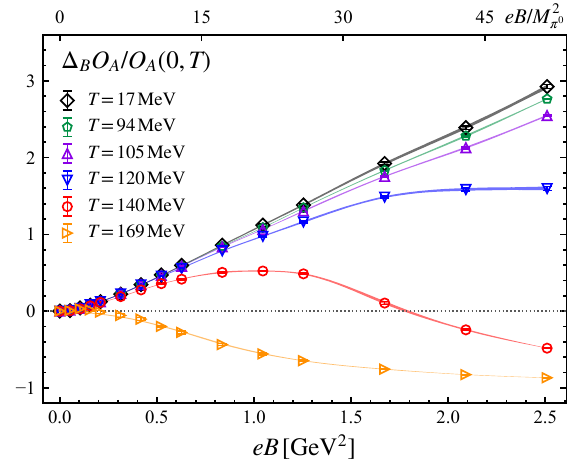}
\caption{Relative magnetic response of the neutral-sector partner splittings at fixed temperature. Left: $R_{O_\chi}(eB,T)=\Delta_B O_\chi(eB,T)/O_\chi(eB=0,T)$. Right: $R_{O_A}(eB,T)=\Delta_B O_A(eB,T)/O_A(eB=0,T)$. Here $\Delta_B O(eB,T)=O(eB,T)-O(eB=0,T)$, and the dotted horizontal line marks zero magnetic response.}
\label{fig:relative_response_results}
\end{figure*}

The middle panel of~\autoref{fig:O_eB_results} shows the singlet $U(1)_A$ partner splitting $O_A$. At the same low temperatures, i.e., $T=17$, $94$, and $105$ MeV, the data points increase monotonically with $eB$ over the magnetic-field range studied. Thus, as in the neutral non-singlet chiral channel, the magnetic field enhances the singlet $U(1)_A$ partner splitting at low temperature. At $T=120$ MeV, $O_A$ continues to increase with $eB$ and tends to saturate at large magnetic field.  This differs from $O_\chi$, which already shows a clear decrease at
the largest magnetic fields at the same temperature. As the temperature is raised to $T=140$ MeV, the magnetic-field dependence of $O_A$ becomes non-monotonic.  The splitting first increases at small and intermediate magnetic fields, reaches a maximum around $eB\simeq 1.1$--$1.3~{\rm GeV}^2$, and then decreases at larger $eB$. Compared with $O_\chi$, however, the decrease of $O_A$ sets in at a somewhat
larger magnetic field and is milder.  At $T=169$ MeV, the singlet $U(1)_A$ partner splitting is already small and is further reduced as $eB$ increases. At $T=211$ and $281$ MeV, $O_A$ is strongly suppressed on the scale of the main plot.

This large-field reduction of $O_A$ should be distinguished from the
usual inverse magnetic catalysis of the chiral condensate.  In the
non-singlet chiral channel, the suppression of $O_\chi$ at sufficiently
large $eB$ is the susceptibility counterpart of the
inverse-magnetic-catalysis pattern already known from the light-quark
condensate.  For $O_A$, the corresponding behavior concerns the
magnetic-field dependence of singlet $U(1)_A$-breaking effects in the
neutral partner splitting.  It is therefore more appropriate to describe
it as an inverse-magnetic-catalysis-like response of the singlet
$U(1)_A$ partner splitting.  A qualitatively similar scenario was
proposed in the model study of Ref.~\cite{Wang:2021dcy}, where it was
referred to as axial inverse magnetic catalysis.  The present result
provides, to our knowledge, the first lattice-QCD determination of this
response from a neutral-sector $U(1)_A$ partner splitting.  The lattice data
indicate that singlet $U(1)_A$-breaking effects, like the non-singlet
chiral splitting, can be reduced by sufficiently strong magnetic fields
in the intermediate-temperature range, although the onset of this
reduction is shifted to larger $eB$ or higher temperature.

The right panel of~\autoref{fig:O_eB_results} shows $O_{\rm disc}=O_A-O_\chi$, which directly measures the difference between the singlet $U(1)_A$ partner splitting and the neutral non-singlet chiral splitting. At low temperatures, $O_{\rm disc}$ is much
smaller than $O_\chi$ and $O_A$, consistent with the similar magnetic enhancement of the two partner splittings in this temperature range.  The difference becomes much more pronounced at $T=120$ and $140$ MeV. At $T=120$ MeV, $O_{\rm disc}$ increases strongly with $eB$ and reaches its largest value at the largest magnetic field shown. At $T=140$ MeV, $O_{\rm disc}$ also increases with $eB$ at small and intermediate magnetic fields, reaches a broad maximum, and then decreases at the largest magnetic fields.  This behavior reflects the fact that $O_\chi$ and $O_A$ no longer
respond to the magnetic field in the same way in this temperature range. At $T=169$ MeV, $O_{\rm disc}$ is smaller and decreases with increasing $eB$, while at $T=211$ and $281$ MeV it remains close to zero on the scale of
the main plot.

To quantify the relative size of the magnetic-field-induced changes already visible in \autoref{fig:O_eB_results}, we consider the normalized magnetic response
\begin{equation}
\begin{split}
    R_{O_i}(eB,T)
    &\equiv 
    \frac{\Delta_B O_i(eB,T)}{O_i(eB=0,T)} \\
    & =
    \frac{O_i(eB,T)-O_i(eB=0,T)}{O_i(eB=0,T)} .
    \label{eq:relative_response_def}
\end{split}
\end{equation}
  The results for
$O_\chi$ and $O_A$ are shown in \autoref{fig:relative_response_results}.  This
figure is therefore a normalized representation of the magnetic response shown in \autoref{fig:O_eB_results}, and is used to compare the fractional response of the neutral non-singlet chiral and singlet $U(1)_A$ partner splittings~\footnote{Since $O_\chi(eB=0,T)$ and $O_A(eB=0,T)$ are already very small at
$T=211$ and $281$ MeV, the corresponding ratios can amplify tiny absolute differences; we therefore use \autoref{fig:relative_response_results} mainly
for the quantitative comparison up to $T=169$ MeV, while the high-temperature suppression is assessed from the observables in \autoref{fig:O_T_results} and \autoref{fig:O_eB_results}.}. The shaded bands in \autoref{fig:relative_response_results} are constructed as in \autoref{fig:O_T_results} and \autoref{fig:O_eB_results}, but using one-dimensional smoothing fits in $eB$ at fixed temperature.

The left panel of~\autoref{fig:relative_response_results} shows $R_{O_\chi}$ for the neutral non-singlet chiral splitting. At low temperatures, $T=17$, $94$, and $105$ MeV, the response is
positive over the full magnetic-field range.  At the largest magnetic field, the ratio reaches values of order $2.5$--$3$, meaning that $O_\chi$ is enhanced
by several hundred percent compared with its value at $eB=0$.  At $T=120$ MeV, the response is still positive, but its growth is reduced at large $eB$; this is the normalized counterpart of the turnover seen in the
left panel of \autoref{fig:O_eB_results}.  At $T=140$ MeV, $R_{O_\chi}$ is positive at small and intermediate magnetic fields, but changes sign at
around $eB\simeq 1.4~{\rm GeV}^2$ and becomes negative at larger fields.  Thus the large-field data at this temperature correspond to a fractional decrease
of $O_\chi$ relative to its zero-field value.  At $T=169$ MeV, the response is negative for most nonzero magnetic fields and approaches $R_{O_\chi}\simeq -1$ at large $eB$, showing that the already small neutral non-singlet chiral splitting is further suppressed by the magnetic field.

The right panel of~\autoref{fig:relative_response_results} shows $R_{O_A}$ for the singlet $U(1)_A$ partner splitting. At low temperatures, $T=17$, $94$, and $105$ MeV, the fractional response is positive over the full magnetic-field range and is close in size to $R_{O_\chi}$.  At the largest magnetic field, both ratios reach values of
order $2.5$--$3$, showing that both partner splittings are enhanced by several hundred percent relative to their zero-field values.  The difference between the two channels becomes more visible at $T=120$ MeV.  At this temperature, $R_{O_A}$ remains positive and reaches about $1.5$--$1.6$ at large $eB$, whereas $R_{O_\chi}$ has already started to decrease from its maximum and is smaller at the largest magnetic fields. Thus, in fractional terms, the magnetic enhancement of the singlet $U(1)_A$ partner splitting persists more strongly than that of the neutral non-singlet chiral splitting at
$T=120$ MeV.

At $T=140$ MeV, both ratios become non-monotonic, but the reduction is much milder for $O_A$.  The neutral non-singlet chiral response $R_{O_\chi}$ changes
sign at around $eB\simeq 1.4~{\rm GeV}^2$ and approaches $-1$ at the largest magnetic fields.  By contrast, $R_{O_A}$ remains positive up to a larger magnetic field, changes sign only around $eB\simeq 1.8~{\rm GeV}^2$, and reaches only about $-0.5$ at the largest $eB$.  In other words, at the largest magnetic field the neutral non-singlet chiral splitting is almost completely suppressed relative to its zero-field value, while the singlet $U(1)_A$ partner splitting is reduced by roughly one half.  At $T=169$ MeV,
$R_{O_A}$ is negative over most of the magnetic-field range, as is $R_{O_\chi}$, but its magnitude remains slightly smaller at large $eB$.  
On a fractional scale, this comparison shows that the magnetic reduction
sets in more mildly for the singlet $U(1)_A$ partner splitting than for
the neutral non-singlet chiral splitting.

%%%%%%%%%%%%%%%%%%%%%%%%%%%%%%%%%%%%%%%%%%%%%%%%%%%%%%%%%%%%
\section{Conclusions}
\label{sec:conclusions}
%%%%%%%%%%%%%%%%%%%%%%%%%%%%%%%%%%%%%%%%%%%%%%%%%%%%%%%%%%%%
We have studied neutral-sector probes of chiral and singlet
$U(1)_A$ symmetries in $(2+1)$-flavor QCD in a background
magnetic field.  Because the light-quark electric charges are different, a magnetic field explicitly reduces the non-singlet flavor symmetry.  We therefore first identified the appropriate neutral partner channels in a pure magnetic background.  The neutral-pion--sigma susceptibility difference defines the
chiral-partner splitting $O_\chi$ associated with the surviving neutral non-singlet axial symmetry $U(1)^{(3)}_A$, while the
neutral-pion--delta susceptibility difference defines the singlet $U(1)_A$ partner splitting $O_A$.  Their difference, $O_{\rm disc}=O_A-O_\chi$, displays the separation between the two partner splittings.

Numerically, we use fixed-scale HISQ ensembles with $m_l=m_s^{\rm phys}/10$, corresponding to $M_\pi\simeq 220$ MeV at $eB=0$. The temperature dependence of the splittings shows that the neutral non-singlet chiral splitting $O_\chi$ is rapidly reduced between $T=120$ and $169$ MeV. Although $O_\chi$ is still nonzero at $T=169$ MeV, it is much smaller than in the low-temperature range and becomes close to zero on the scale of the present data at $T=211$ and $281$ MeV, indicating the approach toward effective restoration of the surviving neutral non-singlet chiral symmetry at high temperature. The singlet $U(1)_A$ partner splitting $O_A$ follows the same overall thermal trend but is more persistent: residual values remain visible around $T=211$ MeV, while at $T=281$ MeV it is strongly suppressed for all magnetic fields studied.

The magnetic-field dependence reveals a strongly temperature-dependent response. At low temperatures, both $O_\chi$ and $O_A$ increase with $eB$. For $O_\chi$, this increase provides the neutral non-singlet chiral-partner susceptibility counterpart of magnetic catalysis known from the light-quark condensate, while for $O_A$ it represents an analogous magnetic-catalysis-like enhancement of the singlet $U(1)_A$ partner splitting. 
In the intermediate-temperature range, sufficiently strong magnetic fields reduce the splittings. The reduction is milder for $O_A$ and sets in at larger magnetic fields than for $O_\chi$, showing that the neutral chiral and singlet axial partner splittings do not respond
identically to the magnetic field. For $O_\chi$, this reduction provides the neutral non-singlet chiral-partner susceptibility counterpart of the inverse-magnetic-catalysis pattern known from the light-quark condensate. For $O_A$, the corresponding behavior concerns the magnetic-field dependence of singlet $U(1)_A$-breaking effects in the neutral partner splitting: the data show an inverse-magnetic-catalysis-like reduction of
the singlet $U(1)_A$ partner splitting. This behavior  qualitatively parallels with the model scenario referred to as axial inverse magnetic
catalysis. The present calculation provides, to our knowledge, the first lattice-QCD determination of susceptibility-splitting counterparts of both magnetic catalysis and inverse magnetic catalysis in the neutral chiral channel, together with the analogous low-temperature enhancement and large-field suppression in the singlet $U(1)_A$ channel.
 
The present study is performed at a single lattice spacing and at a heavier-than-physical pion mass.  The results should therefore be regarded as fixed-scale lattice-QCD evidence for the magnetic response of the neutral partner splittings.  Future calculations at smaller lattice spacings and lighter pion masses will be needed to quantify the continuum and chiral behavior, and to further connect the observed singlet $U(1)_A$ response with topology and the infrared Dirac spectrum in magnetic fields.

\section*{Acknowledgments}

This work is supported partly by the National Natural Science Foundation of China under Grants No. 12293064, No. 12293060, and No. 12325508, as well as the National Key Research and Development Program of China under Contract No. 2022YFA1604900 and the Fundamental Research Funds for the Central Universities, Central China Normal University under Grants No. 30101250314 and No.30106250152. The numerical simulations have been performed on the GPU cluster in the Nuclear Science Computing Center at Central China Normal University ($\mathrm{NSC}^{3}$) and Wuhan Supercomputing Center.

\bibliographystyle{JHEP.bst}
\bibliography{refs.bib}

\providecommand{\href}[2]{#2}\begingroup\raggedright\begin{thebibliography}{10}

\bibitem{Adler:1969gk}
S.L.~Adler, \emph{{Axial vector vertex in spinor electrodynamics}},
  \href{https://doi.org/10.1103/PhysRev.177.2426}{\emph{Phys. Rev.} {\bfseries
  177} (1969) 2426}.

\bibitem{Bell:1969ts}
J.S.~Bell and R.~Jackiw, \emph{{A PCAC puzzle: $\pi^0 \to \gamma \gamma$ in the
  $\sigma$ model}}, \href{https://doi.org/10.1007/BF02823296}{\emph{Nuovo Cim.
  A} {\bfseries 60} (1969) 47}.

\bibitem{tHooft:1976rip}
G.~'t~Hooft, \emph{{Symmetry Breaking Through Bell-Jackiw Anomalies}},
  \href{https://doi.org/10.1103/PhysRevLett.37.8}{\emph{Phys. Rev. Lett.}
  {\bfseries 37} (1976) 8}.

\bibitem{Cohen:1996ng}
T.D.~Cohen, \emph{{The High temperature phase of QCD and U(1)-A symmetry}},
  \href{https://doi.org/10.1103/PhysRevD.54.R1867}{\emph{Phys.Rev.} {\bfseries
  D54} (1996) 1867} [\href{https://arxiv.org/abs/hep-ph/9601216}{{\ttfamily
  hep-ph/9601216}}].

\bibitem{Lee:1996zy}
S.H.~Lee and T.~Hatsuda, \emph{{U-a(1) symmetry restoration in QCD with N(f)
  flavors}}, \href{https://doi.org/10.1103/PhysRevD.54.R1871}{\emph{Phys. Rev.
  D} {\bfseries 54} (1996) R1871}
  [\href{https://arxiv.org/abs/hep-ph/9601373}{{\ttfamily hep-ph/9601373}}].

\bibitem{Shuryak:1993ee}
E.V.~Shuryak, \emph{{Which chiral symmetry is restored in hot QCD?}},
  {\emph{Comments Nucl. Part. Phys.} {\bfseries 21} (1994) 235}
  [\href{https://arxiv.org/abs/hep-ph/9310253}{{\ttfamily hep-ph/9310253}}].

\bibitem{Pisarski:1983ms}
R.D.~Pisarski and F.~Wilczek, \emph{{Remarks on the Chiral Phase Transition in
  Chromodynamics}},
  \href{https://doi.org/10.1103/PhysRevD.29.338}{\emph{Phys.Rev.} {\bfseries
  D29} (1984) 338}.

\bibitem{Banks:1979yr}
T.~Banks and A.~Casher, \emph{{Chiral Symmetry Breaking in Confining
  Theories}}, \href{https://doi.org/10.1016/0550-3213(80)90255-2}{\emph{Nucl.
  Phys. B} {\bfseries 169} (1980) 103}.

\bibitem{Ding:2023oxy}
H.-T.~Ding, W.-P.~Huang, S.~Mukherjee and P.~Petreczky, \emph{{Microscopic
  Encoding of Macroscopic Universality: Scaling Properties of Dirac
  Eigenspectra near QCD Chiral Phase Transition}},
  \href{https://doi.org/10.1103/PhysRevLett.131.161903}{\emph{Phys. Rev. Lett.}
  {\bfseries 131} (2023) 161903}
  [\href{https://arxiv.org/abs/2305.10916}{{\ttfamily 2305.10916}}].

\bibitem{Leutwyler:1992yt}
H.~Leutwyler and A.V.~Smilga, \emph{{Spectrum of Dirac operator and role of
  winding number in QCD}},
  \href{https://doi.org/10.1103/PhysRevD.46.5607}{\emph{Phys. Rev. D}
  {\bfseries 46} (1992) 5607}.

\bibitem{Evans:1996wf}
N.J.~Evans, S.D.H.~Hsu and M.~Schwetz, \emph{{Topological charge and U(1)-A
  symmetry in the high temperature phase of QCD}},
  \href{https://doi.org/10.1016/0370-2693(96)00280-8}{\emph{Phys. Lett. B}
  {\bfseries 375} (1996) 262}
  [\href{https://arxiv.org/abs/hep-ph/9601361}{{\ttfamily hep-ph/9601361}}].

\bibitem{Ding:2020xlj}
H.T.~Ding, S.T.~Li, S.~Mukherjee, A.~Tomiya, X.D.~Wang and Y.~Zhang,
  \emph{{Correlated Dirac eigenvalues and axial anomaly in chiral symmetric
  QCD}}, \href{https://doi.org/10.1103/PhysRevLett.126.082001}{\emph{Phys. Rev.
  Lett.} {\bfseries 126} (2021) 082001}
  [\href{https://arxiv.org/abs/2010.14836}{{\ttfamily 2010.14836}}].

\bibitem{Aoki:2012yj}
S.~Aoki, H.~Fukaya and Y.~Taniguchi, \emph{{Chiral symmetry restoration,
  eigenvalue density of Dirac operator and axial U(1) anomaly at finite
  temperature}},
  \href{https://doi.org/10.1103/PhysRevD.86.114512}{\emph{Phys.Rev.} {\bfseries
  D86} (2012) 114512} [\href{https://arxiv.org/abs/1209.2061}{{\ttfamily
  1209.2061}}].

\bibitem{HotQCD:2012vvd}
{\scshape HotQCD} collaboration, \emph{{The chiral transition and $U(1)_A$
  symmetry restoration from lattice QCD using Domain Wall Fermions}},
  \href{https://doi.org/10.1103/PhysRevD.86.094503}{\emph{Phys. Rev. D}
  {\bfseries 86} (2012) 094503}
  [\href{https://arxiv.org/abs/1205.3535}{{\ttfamily 1205.3535}}].

\bibitem{Buchoff:2013nra}
M.I.~Buchoff, M.~Cheng, N.H.~Christ, H.T.~Ding, C.~Jung et~al., \emph{{QCD
  chiral transition, U(1)A symmetry and the dirac spectrum using domain wall
  fermions}},
  \href{https://doi.org/10.1103/PhysRevD.89.054514}{\emph{Phys.Rev.} {\bfseries
  D89} (2014) 054514} [\href{https://arxiv.org/abs/1309.4149}{{\ttfamily
  1309.4149}}].

\bibitem{Brandt:2024gso}
B.B.~Brandt, G.~Endr{\H{o}}di, J.J.H.~Hern{\'a}ndez and G.~Mark{\'o},
  \emph{{Impact of extreme magnetic fields on the QCD topological
  susceptibility in the vicinity of the crossover region}},
  \href{https://doi.org/10.1007/JHEP12(2024)228}{\emph{JHEP} {\bfseries 12}
  (2025) 228} [\href{https://arxiv.org/abs/2409.00796}{{\ttfamily
  2409.00796}}].

\bibitem{Adhikari:2021lbl}
P.~Adhikari, \emph{{Topological susceptibility in a uniform magnetic field}},
  \href{https://doi.org/10.1016/j.physletb.2021.136826}{\emph{Phys. Lett. B}
  {\bfseries 825} (2022) 136826}
  [\href{https://arxiv.org/abs/2103.05048}{{\ttfamily 2103.05048}}].

\bibitem{Kharzeev:2007jp}
D.E.~Kharzeev, L.D.~McLerran and H.J.~Warringa, \emph{{The Effects of
  topological charge change in heavy ion collisions: 'Event by event P and CP
  violation'}},
  \href{https://doi.org/10.1016/j.nuclphysa.2008.02.298}{\emph{Nucl. Phys.}
  {\bfseries A803} (2008) 227}
  [\href{https://arxiv.org/abs/0711.0950}{{\ttfamily 0711.0950}}].

\bibitem{Fukushima:2008xe}
K.~Fukushima, D.E.~Kharzeev and H.J.~Warringa, \emph{{The Chiral Magnetic
  Effect}}, \href{https://doi.org/10.1103/PhysRevD.78.074033}{\emph{Phys. Rev.
  D} {\bfseries 78} (2008) 074033}
  [\href{https://arxiv.org/abs/0808.3382}{{\ttfamily 0808.3382}}].

\bibitem{Kharzeev:2015znc}
D.E.~Kharzeev, J.~Liao, S.A.~Voloshin and G.~Wang, \emph{{Chiral magnetic and
  vortical effects in high-energy nuclear collisions---A status report}},
  \href{https://doi.org/10.1016/j.ppnp.2016.01.001}{\emph{Prog. Part. Nucl.
  Phys.} {\bfseries 88} (2016) 1}
  [\href{https://arxiv.org/abs/1511.04050}{{\ttfamily 1511.04050}}].

\bibitem{Yamamoto:2021oys}
A.~Yamamoto, \emph{{Overview of external electromagnetism and rotation in
  lattice QCD}},
  \href{https://doi.org/10.1140/epja/s10050-021-00530-8}{\emph{Eur. Phys. J. A}
  {\bfseries 57} (2021) 211}
  [\href{https://arxiv.org/abs/2103.00237}{{\ttfamily 2103.00237}}].

\bibitem{Endrodi:2024cqn}
G.~Endrodi, \emph{{QCD with background electromagnetic fields on the lattice: A
  review}}, \href{https://doi.org/10.1016/j.ppnp.2024.104153}{\emph{Prog. Part.
  Nucl. Phys.} {\bfseries 141} (2025) 104153}
  [\href{https://arxiv.org/abs/2406.19780}{{\ttfamily 2406.19780}}].

\bibitem{Brandt:2026dcd}
B.B.~Brandt and G.~Endrodi, \emph{{Thermodynamics of magnetized matter in hot
  and dense QCD}},  \href{https://arxiv.org/abs/2604.26715}{{\ttfamily
  2604.26715}}.

\bibitem{Ding:2026gao}
H.-T.~Ding, \emph{{Lattice QCD at finite temperature and density}},  in
  \emph{{42th International Symposium on Lattice Field Theory}}, 3, 2026
  [\href{https://arxiv.org/abs/2603.16230}{{\ttfamily 2603.16230}}].

\bibitem{DElia:2011koc}
M.~D'Elia and F.~Negro, \emph{{Chiral Properties of Strong Interactions in a
  Magnetic Background}},
  \href{https://doi.org/10.1103/PhysRevD.83.114028}{\emph{Phys. Rev. D}
  {\bfseries 83} (2011) 114028}
  [\href{https://arxiv.org/abs/1103.2080}{{\ttfamily 1103.2080}}].

\bibitem{Bali:2011qj}
G.S.~Bali, F.~Bruckmann, G.~Endrodi, Z.~Fodor, S.D.~Katz, S.~Krieg et~al.,
  \emph{{The QCD phase diagram for external magnetic fields}},
  \href{https://doi.org/10.1007/JHEP02(2012)044}{\emph{JHEP} {\bfseries 02}
  (2012) 044} [\href{https://arxiv.org/abs/1111.4956}{{\ttfamily 1111.4956}}].

\bibitem{Bali:2012zg}
G.S.~Bali, F.~Bruckmann, G.~Endrodi, Z.~Fodor, S.D.~Katz and A.~Schafer,
  \emph{{QCD quark condensate in external magnetic fields}},
  \href{https://doi.org/10.1103/PhysRevD.86.071502}{\emph{Phys. Rev.}
  {\bfseries D86} (2012) 071502}
  [\href{https://arxiv.org/abs/1206.4205}{{\ttfamily 1206.4205}}].

\bibitem{Bruckmann:2013oba}
F.~Bruckmann, G.~Endrodi and T.G.~Kovacs, \emph{{Inverse magnetic catalysis and
  the Polyakov loop}},
  \href{https://doi.org/10.1007/JHEP04(2013)112}{\emph{JHEP} {\bfseries 04}
  (2013) 112} [\href{https://arxiv.org/abs/1303.3972}{{\ttfamily 1303.3972}}].

\bibitem{DElia:2018xwo}
M.~D'Elia, F.~Manigrasso, F.~Negro and F.~Sanfilippo, \emph{{QCD phase diagram
  in a magnetic background for different values of the pion mass}},
  \href{https://doi.org/10.1103/PhysRevD.98.054509}{\emph{Phys. Rev.}
  {\bfseries D98} (2018) 054509}
  [\href{https://arxiv.org/abs/1808.07008}{{\ttfamily 1808.07008}}].

\bibitem{Ding:2022tqn}
H.T.~Ding, S.T.~Li, J.H.~Liu and X.D.~Wang, \emph{{Chiral condensates and
  screening masses of neutral pseudoscalar mesons in thermomagnetic QCD
  medium}}, \href{https://doi.org/10.1103/PhysRevD.105.034514}{\emph{Phys. Rev.
  D} {\bfseries 105} (2022) 034514}
  [\href{https://arxiv.org/abs/2201.02349}{{\ttfamily 2201.02349}}].

\bibitem{Ding:2026qzu}
H.-T.~Ding and D.~Zhang, \emph{{Chiral properties of (2+1)-flavor QCD in
  magnetic fields at zero temperature}},
  \href{https://doi.org/10.1103/qpxq-xjqc}{\emph{Phys. Rev. D} {\bfseries 113}
  (2026) 094503} [\href{https://arxiv.org/abs/2601.18354}{{\ttfamily
  2601.18354}}].

\bibitem{Wang:2021dcy}
Y.~Wang and S.~Matsuzaki, \emph{{Axial inverse magnetic catalysis}},
  \href{https://doi.org/10.1103/PhysRevD.105.074015}{\emph{Phys. Rev. D}
  {\bfseries 105} (2022) 074015}
  [\href{https://arxiv.org/abs/2110.10432}{{\ttfamily 2110.10432}}].

\bibitem{Fujikawa:1979ay}
K.~Fujikawa, \emph{{Path Integral Measure for Gauge Invariant Fermion
  Theories}}, \href{https://doi.org/10.1103/PhysRevLett.42.1195}{\emph{Phys.
  Rev. Lett.} {\bfseries 42} (1979) 1195}.

\bibitem{Kilcup:1986dg}
G.W.~Kilcup and S.R.~Sharpe, \emph{{A Tool Kit for Staggered Fermions}},
  \href{https://doi.org/10.1016/0550-3213(87)90285-9}{\emph{Nucl. Phys.}
  {\bfseries B283} (1987) 493}.

\bibitem{Gregory:2007ev}
E.B.~Gregory, A.C.~Irving, C.M.~Richards and C.~McNeile, \emph{{Methods for
  Pseudoscalar Flavour-Singlet Mesons with Staggered Fermions}},
  \href{https://doi.org/10.1103/PhysRevD.77.065019}{\emph{Phys. Rev. D}
  {\bfseries 77} (2008) 065019}
  [\href{https://arxiv.org/abs/0709.4224}{{\ttfamily 0709.4224}}].

\bibitem{Donald:2011if}
G.C.~Donald, C.T.H.~Davies, E.~Follana and A.S.~Kronfeld, \emph{{Staggered
  fermions, zero modes, and flavor-singlet mesons}},
  \href{https://doi.org/10.1103/PhysRevD.84.054504}{\emph{Phys. Rev. D}
  {\bfseries 84} (2011) 054504}
  [\href{https://arxiv.org/abs/1106.2412}{{\ttfamily 1106.2412}}].

\bibitem{Ding:2020hxw}
H.T.~Ding, S.T.~Li, A.~Tomiya, X.D.~Wang and Y.~Zhang, \emph{{Chiral properties
  of (2+1)-flavor QCD in strong magnetic fields at zero temperature}},
  \href{https://doi.org/10.1103/PhysRevD.104.014505}{\emph{Phys. Rev. D}
  {\bfseries 104} (2021) 014505}
  [\href{https://arxiv.org/abs/2008.00493}{{\ttfamily 2008.00493}}].

\bibitem{Ding:2021cwv}
H.T.~Ding, S.T.~Li, Q.~Shi and X.D.~Wang, \emph{{Fluctuations and correlations
  of net baryon number, electric charge and strangeness in a background
  magnetic field}},
  \href{https://doi.org/10.1140/epja/s10050-021-00519-3}{\emph{Eur. Phys. J. A}
  {\bfseries 57} (2021) 202}
  [\href{https://arxiv.org/abs/2104.06843}{{\ttfamily 2104.06843}}].

\end{thebibliography}\endgroup

%%%%%%%%%%%%%%%%%%%%%%%%%%%%%%%%%%%%%%%%%%%%%%%%%%%%%%%%%%%%
\appendix
%%%%%%%%%%%%%%%%%%%%%%%%%%%%%%%%%%%%%%%%%%%%%%%%%%%%%%%%%%%%

%%%%%%%%%%%%%%%%%%%%%%%%%%%%%%%%%%%%%%%%%%%%%%%%%%%%%%%%%%%%
\section{Ward-identity check for the neutral-pion susceptibility}
\label{app:pi0_WI_check}
%%%%%%%%%%%%%%%%%%%%%%%%%%%%%%%%%%%%%%%%%%%%%%%%%%%%%%%%%%%%

The full neutral-pion susceptibility is obtained from the Ward identity
in \autoref{eq:pi0_ward_identity}.  As a consistency check, we compare
it with the connected pion susceptibility by considering
\begin{equation}
  \mathcal{R}_{\pi^0}
  =
  \frac{
  \langle\bar u u\rangle+\langle\bar d d\rangle
  }{
  2m_l\,\chi_{5,\rm con}
  } .
\label{eq:R_pi0_WI}
\end{equation}

\begin{figure}[ht!]
  \centering
  \includegraphics[width=0.4\textwidth]{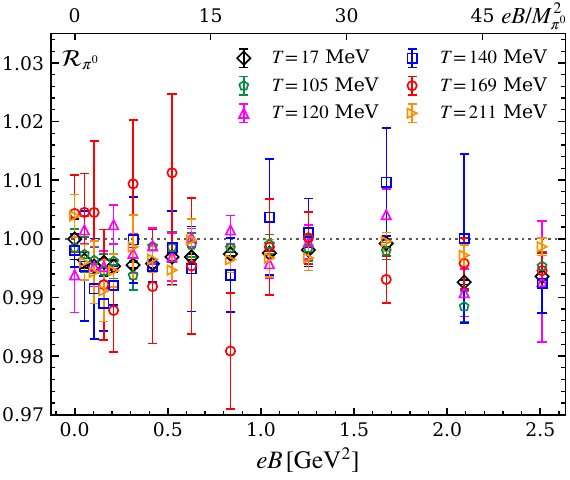}
  \caption{
  Integrated Ward-identity check for the neutral-pion susceptibility.
  The ratio $\mathcal{R}_{\pi^0}$ defined in \autoref{eq:R_pi0_WI} is shown as a
  function of $eB$ at representative temperatures.  The connected pion
  susceptibility in the denominator is obtained from the available
  temporal and/or spatial correlator sums; when both determinations are
  available, they are averaged.  Errors are estimated by a block
  bootstrap analysis.  The dotted line marks $\mathcal{R}_{\pi^0}=1$.
  }
  \label{fig:R_pi0_WI_check}
\end{figure}

In the continuum pure-$B$ theory, the integrated minus pseudoscalar
disconnected susceptibility vanishes, and hence
$\chi_{\pi^0}=\chi_{5,\rm con}$. Therefore $\mathcal{R}_{\pi^0}=1$ is expected up
to finite-lattice-spacing effects and statistical uncertainties.  The
deviation of $\mathcal{R}_{\pi^0}$ from unity provides a practical estimate of the
size of these effects for $\chi^{(-)}_{5,\rm disc}$ in the present calculation. 
We plot $\mathcal{R}_{\pi^0}$ in Fig.~\ref{fig:R_pi0_WI_check} as a function of the magnetic field at selected temperatures to avoid over-cluttering. As can be seen in the image, $\mathcal{R}_{\pi^0}\approx1$ within 2\% for all studied temperatures and magnetic fields, demonstrating that indeed $\chi^{(-)}_{5,\rm disc}\approx0$ even for our current lattice setup.

\end{document}